\documentclass[11pt]{article}
\usepackage{amsthm,amsfonts,amssymb,amsmath}

\makeatletter

\@addtoreset{equation}{section}

\makeatother

\newcommand{\p}{\partial}
\newcommand{\wt}{\widetilde}
\newcommand{\comp}{\mathbb C}
\newcommand{\inte}{\mathbb Z}

\newcommand{\cL}{{\cal L}}
\newcommand{\cA}{{\cal A}}
\newcommand{\cB}{{\cal B}}
\newcommand{\cE}{{\cal E}}
\newcommand{\vL}{\Vec{\Lambda}}
\newcommand{\vB}{\Vec{B}}
\newcommand{\vS}{\Vec{S}}
\newcommand{\vU}{\Vec{U}}
\newcommand{\vV}{\Vec{V}}
\newcommand{\vZ}{\Vec{Z}}

\DeclareMathOperator{\Tr}{Tr}
\newcommand{\res}{\mathop{\rm res}\nolimits}

\theoremstyle{plain}
\newtheorem{teo}{Theorem}
\newtheorem{lem}{Lemma}

\theoremstyle{remark}
\newtheorem*{rem}{Remark}

\begin{document}

\begin{titlepage}

\title{Elliptic Families of Solutions of the
Kadomtsev\,--\,Petviashvili Equation and
the Field Elliptic Calogero\,--\,Moser
System\thanks{Research
is supported in part by the National Science
Foundation under the grant DMS-01-04621.}}

\bigskip
\author{A. Akhmetshin${}^{**}$\and I. Krichever${}^{\dagger}$\and
Yu. Volvovski${}^{\ddagger}$}
\date{}
\maketitle

\thispagestyle{empty}

\begin{center}
\small
${}^{**\,\ddagger}$Department of Mathematics,\\
Columbia University,\\
2990 Broadway, Mail Code 4406, New York, NY 10027, USA
\end{center}

\begin{center}
\small
${}^{\dagger}$Department of Mathematics,\\
Columbia University,\\
and\\
Landau Institute for Theoretical Physics,\\
Kosygina 2, 117334 Moscow, Russia,\\
and\\
Institute for Theoretical and Experimental Physics,\\
B.~Cheremushkinskaja 25, 117259 Moscow, Russia
\end{center}

\begin{center}
\small
e-mail:\\
${}^{**}$\verb"alakhm@math.columbia.edu"\\
${}^{\dagger}$\verb"krichev@math.columbia.edu"\\
${}^{\ddagger}$\verb"yurik@math.columbia.edu"
\end{center}

\bigskip
\begin{abstract}
We present the Lax pair for the field elliptic Calogero\,--\,Moser
system and establish a connection between this system and the
Kadomtsev\,--\,Petviashvili equation.
Namely, we consider elliptic families of solutions of the KP equation,
such that their poles satisfy a constraint of being \emph{balanced}.
We show that the dynamics of these poles is described by a reduction
of the field elliptic CM system.

We construct a wide class of solutions to the field elliptic
CM system by showing that any $N$-fold branched cover of an
elliptic curve gives rise to an elliptic family of solutions
of the KP equation with balanced poles.
\end{abstract}

\vfill

\end{titlepage}

\section{Introduction}

The main goal of this paper is to establish a connection between
the field analog of the elliptic Calogero\,--\,Moser system (CM)
introduced in \cite{krvb} and the Kadomtsev\,--\,Petviashvili
equation (KP).
This connection is a next step along the line which goes back to the
work \cite{amkm} where it was found that dynamic of poles of the
elliptic (rational or trigonometric) solutions of the Korteweg\,--\,de
Vries equation (KdV) can be described in terms of commuting flows of
the elliptic (rational or trigonometric) CM system.

In the earlier work of one of the authors \cite{krelkp} it was shown that
constrained correspondence between a theory of the elliptic CM system
and a theory of the elliptic solutions of the KdV equation becomes an
isomorphism for the case of the KP equation. It turns out that a
function $u(x,y,t)$ which is an elliptic function with respect to the
variable~$x$ satisfies the KP equation if and only if it has the form
\begin{equation}\label{pot}
u(x,y,t)=-2\sum_{i=1}^N\wp(x-q_i(y,t))+c,
\end{equation}
and its poles $q_i$ as functions of $y$ satisfy the equations of
motion of the elliptic CM system. The latter is a system of $N$ particles
on an elliptic curve with pairwise interactions. Its Hamiltonian has the
form
$$
H_2=\frac{1}{2}\sum_{i=1}^N p_i^2 - 2 \sum_{i\ne j}\wp(q_i-q_j),
$$
where $\wp(q)$ is the Weierstrass $\wp$-function. The dynamics of
the particles~$q_i$ with respect to~$t$ coincides with
the commuting flow generated by the third Hamiltonian~$H_3$ of the system.
Recall, that the elliptic~CM system is a completely integrable system.
It admits the Lax representation $\dot L=[L,M]$, where~$L=L(z)$ and~$M=M(z)$
are $(N\times N)$ matrices depending on a spectral parameter~$z$ \cite{c}.
The involutive integrals~$H_n$ are defined as $H_n=n^{-1}\Tr L^n$.

An explicit theta-functional formula for algebro-geometric solutions
of the KP equation provides an \emph{algebraic} solution
of the Cauchy problem for the elliptic CM system \cite{krelkp}.
Namely, the positions $q_i(y)$ of the particles at any time~$y$ are roots of
the equation
$$
\theta \bigl(\vU q_i+\vV y+\vZ\,|\,B)=0 ,
$$
where the theta-function $\theta(z\,|\,B)$ is the Riemann theta-function
constructed with the help of the matrix of $b$-periods of the holomorphic
differentials on a \emph{time-independent} spectral curve~$\Gamma$.
The spectral curve is given by $R(k,z)=\det (kI-L(z))=0$
and the vectors $\vU$, $\vV$, $\vZ$ are defined by the initial data.

The correspondence between finite-dimensional integrable systems and
poles systems of various soliton equation has been extensively studied
in \cite{krbab,krnest,kreltoda,krwz,krzab}. A general scheme of
constructing such systems using a specific inverse problem for
linear equations with elliptic coefficents is presented in \cite{krnest}.

A problem we address in this paper is as follows. The KP equation
\begin{equation}\label{kp}
\frac{3}{4} u_{yy}=\frac{\p}{\p x} \biggl( u_t-\frac{1}{4}u_{xxx}-
\frac{3}{2} u u_x \biggr)
\end{equation}
is the first equation of a hierarchy of commuting flows.
A general solution of the whole hierarchy is known to be of
the form
$$
u(x,y,t,t_4\ldots)=2 \frac{\p^2}{\p x^2}\ln\tau(x,y,t,t_4\ldots),
\qquad x=t_1,\ y=t_2,\ t=t_3,
$$
where~$\tau$ is the so-called KP \emph{tau}-function.
We consider solutions $u$ that are elliptic function with
respect to some variable $t_k$ or a linear combination of times
$\lambda=\sum_k\alpha_kt_k$.

It is instructive to consider first the algebraic-geometrical solutions of
the KP equation. According to \cite{kr} any smooth algebraic curve~$\Gamma$
with a puncture defines a solution of the KP hierarchy by the formula
\begin{equation}
u=2 \frac{\p^2}{\p x^2}\ln \theta \left(\bigl.
\sum\nolimits_{k} \vU_k t_k+\vZ\, \bigr|\,B \right),\qquad x=t_1,
\end{equation}
where as before $B$ is the matrix of $b$-periods of the normalized holomorphic
differentials on~$\Gamma$ and~$\vZ$ is the vector of Riemann constants.
The vectors~$\vU_k$ are the vectors of $b$-periods of
certain meromorphic differentials on~$\Gamma$. The algebraic-geometrical
solution is elliptic with respect to some direction if there is a
vector~$\vL$ which spans an elliptic curve~$\cE$ embedded in the
Jacobian~$J(\Gamma)$. This is a nontrivial constraint and the space of
corresponding algebraic curves has codimension $g-1$ in the moduli space
of all the curves. If the vector~$\vL$ does exist then the theta-divisor
intersects the shifted elliptic curve $\cE+\sum_{k}\vU_kt_k$ at
a finite number of points $\lambda_i(t_1,t_2, \ldots)$.

It can be shown directly that if $u(x,y,t,\lambda)$ is an elliptic
family of solutions of the KP equation, then it has the form
\begin{equation}\label{ukp}
u=-2 \sum_{i=1}^N \bigl[ \lambda_{i\,x}^2\wp(\lambda-\lambda_i) -
\lambda_{i\,xx}\zeta(\lambda-\lambda_i) \bigr] +
c(x,y,t),\qquad \lambda_i=\lambda_i(x,y,t).
\end{equation}
The sum of the residues vanishes for an elliptic function~$u$.
Therefore, $\sum_i\lambda_{ixx}=0$.
We shall consider only solutions~$u$ with the poles~$\lambda_i$
satisfying an additional constraint.
Namely, we say that the poles $\lambda_i$, $i=1,\ldots,N$
are \emph{balanced} if they can be presented in the form
\begin{equation}\label{bal}
\lambda_i(x,y,t)=q_i(x,y,t)-h x,\qquad
\sum_{i=1}^N q_i(x,y,t)=const ,
\end{equation}
where $h$ is an arbitrary non-zero constant.
We prove that if the poles of $u$ are balanced, then the functions~$q_i(x,y)$
satisfy the following equations:
\begin{equation}\label{qyy}
\begin{aligned}
q_{i\,yy} = &-\left\{ \frac{q_{i\,y}^2}{h-q_{i\,x}} \right\}_{\! x}
 +\frac{1}{Nh}(h-q_{i\,x}) \sum_{k=1}^N \left\{ \frac{q_{k\,y}^2}{h-q_{k\,x}}
 \right\}_{\! x}+{}\\
&{}+2(h-q_{i\,x})\frac{\delta U(q)}{\delta q_i}-\frac{2}{Nh}(h-q_{i\,x})
\sum_{k=1}^N (h-q_{k\,x})\frac{\delta U(q)}{\delta q_k} ,
\qquad 1\le i\le N,
\end{aligned}
\end{equation}
where
\begin{equation}\label{U}
\begin{aligned}
U(q)=\sum_{i=1}^N \frac{q^2_{i\,xx}}{4(h-q_{i\,x})}
&-\frac{1}{2} \sum_{j\neq i} \bigl[ (h-q_{j\,x})q_{i\,xx}-(h-q_{i\,x})q_{j\,xx} \bigr]
\zeta(q_i-q_j)+{}\\
{}&+\frac{1}{2}\sum_{j\neq i}
\bigl[(h-q_{j\,x})^2(h-q_{i\,x})+(h-q_{j\,x})(h-q_{i\,x})^2 \bigr]\wp(q_i-q_j) .
\end{aligned}
\end{equation}
Here $\delta/\delta q_i$ is the variational derivative. Since $U(q)$ depends
only on $q_i$ and its two derivatives, we have
$$
\frac{\delta U(q)}{\delta q_i}=\frac{\p U(q)}{\p q_i}-
\frac{d}{dx}\frac{\p U(q)}{\p q_{i\,x}}+
\frac{d^2}{d x^2}\frac{\p U(q)}{\p q_{i\,xx}}, \qquad 1\le i\le N,
$$
Equations (\ref{qyy}) can be identified with a reduction of a particular
case of the Hamiltonian system introduced in \cite{krvb}. We call the
latter system a field analog of the elliptic Calogero\,--\,Moser sytem.
The phase space for this system is the space of functions
$q_1(x),\dots,q_N(x)$, $p_1(x),\dots,p_N(x)$, the Poisson brackets are given by
$$
\{q_i(x),p_j(\tilde x)\}=\delta_{ij}\delta(x-\tilde x) .
$$
and the Hamiltonian equals
\begin{equation}\label{ham}
\widehat H=\int H(x)\,dx, \qquad
H= \sum_{i=1}^N p_i^2(h-q_{i\,x})-
\frac{1}{N h} \biggl( \sum_{i=1}^N p_i(h-q_{i\,x}) \biggr)^{\!2}-{\wt U}(q),
\end{equation}
where
$$
{\wt U}(q)=U(q)+\frac{\p}{\p x} \biggl(\frac{h}{2} \sum_{i\ne j}
(q_{i\,x}-q_{j\,x}) \zeta(q_i-q_j) \biggr) .
$$
The corresponding equations of motion are presented in section~3,
see (\ref{sys}).
Note, that if $q_i$ do not depend on $x$, then (\ref{ham})
reduces to the Hamiltonian of the elliptic CM system.

In particular, for $N=2$ the hamiltonian reduction of this system
corresponding to the constraint $\sum_{i} q_i=0$ is a hamiltonian system
on the space of two functions $q(x)$, $p(x)$, where we set
$$
q=q_1=-q_2,\qquad \frac{1}{h}p(h^2-q_x^2)=p_1(h-q_x)=-p_2(h-q_x),
$$
The Poisson brackets are canonical, i.e.
$\{q(x),p(\tilde x)\}=\delta(x-\tilde x)$,
while the Hamiltonian density~$H$ in the coordinates $\{p,q\}$ may be
rewritten as
$$
H=\frac{2}{h}p^2(h^2-q_x^2)-h\frac{q^2_{xx}}{2(h^2-q_x^2)}-
2h(h^2-3q_x^2)\wp(2q).
$$
It was noticed by A.\,Shabat that the equations of motion given by
this Hamiltonian are equivalent to Landau\,--\,Lifshitz equation.
This case $N=2$ was independently studied in \cite{loz}.

The paper is organized as follows.
In sections 2 and 3 we show that the field analog of the elliptic CM system
describes a solution of the inverse Picard type problem for the linear equation
\begin{equation}\label{pic}
\left(\frac{\p}{\p y}-\cL\right)\psi(x,y,\lambda)=0, \qquad
\cL=\frac{\p^2}{\p x^2}+u(x,y,\lambda),
\end{equation}
which is one of the equations in the auxiliary linear problem for
the KP equation.  Namely, it turns out that if equation (\ref{pic}) with
a family of elliptic in~$\lambda$ potentials of the form (\ref{ukp})
has~$N$ linearly independent meromorphic in~$\lambda$ {\it double-Bloch}
solutions, then the variables $q_i=\lambda_i+h x$ satisfy the equations
of motion generated by the Hamiltonian (\ref{ham}).
As in \cite{krelkp} this inverse problem provides the Lax representation
for the hamiltonian system (\ref{sys}).

In section 4 we show that if $u(x,y,t,\lambda)$ is an elliptic family of
solutions of the KP equation with balanced poles, then the corresponding
family of operators $\p/\p y-\cL$ has infinitely many double-Bloch solution.
Consequently, the dynamics of $q_i(x,y,t)$ with respect to $y$
coincides with the equations of motion of the field elliptic CM system.
We are quite sure that the dynamics of~$q_i$ with respect to all the times
of the KP hierarchy coincides with the hierarchy of commuting flows
for the system (\ref{sys}), but up to now this question remains open.
We plan to investigate it elsewhere.

In the last section we consider the finite-gap solutions of the KP
hierarchy corresponding to an algebraic curve which is an $N$-fold
branched cover of the elliptic curve. We show that they are elliptic with
respect to a certain linear combination~$\lambda$ of the times~$t_k$.
Moreover, as a function of~$\lambda$ these solutions have precisely~$N$
poles.  Therefore, they provide a wide class of exact algebraic
solutions of the field elliptic CM system.

The definitions and properties of classical elliptic functions
and the Riemann $\theta$-function are gathered in the appendix.

\section{Generating problem}

Let us choose a pair of periods $2\omega_1, 2\omega_2\in\comp$,
where $\Im(\omega_2/\omega_1)>0$.
A meromorphic function $f(\lambda)$ is called \emph{double-Bloch}
if it satisfies the following monodromy properties:
$$
f(\lambda+2\omega_a)=B_a f(\lambda),\qquad a=1,2.
$$
The complex constants $B_a$ are called \emph{Bloch multipliers}.
Equivalently, $f(\lambda)$ is a section of a linear bundle over
the elliptic curve $\cE=\comp/\inte[2\omega_1,2\omega_2]$.

We consider the non-stationary Schr\"odinger operator
$$
\p_y-\cL=\p_y-\p^2_{xx}-u(x,y,\lambda),
\qquad \p_x=\p/\p x, \quad \p_y=\p/\p y ,
$$
where the potential $u(x,y,\lambda)$ is a double-periodic function of
the variable $\lambda$. We do not assume any special dependence
with respect to the other variables. Our goal is to find the
potentials $u(x,y,\lambda)$ such that the equation
\begin{equation}\label{sch}
\left(\p_y-\cL \right) \psi(x,y,\lambda) = 0
\end{equation}
has \emph{sufficiently many} double-Bloch solutions.
The existence of such solutions turns out to be a very
restrictive condition (see the discussion in \cite{krnest}).

The basis in the space of the double-Bloch functions can be written
in terms of the fundamental function $\Phi(\lambda,z)$ defined by
the formula
\begin{equation}\label{phi}
\Phi(\lambda,z)=\frac{\sigma(z-\lambda)}{\sigma(z)\sigma(\lambda)}
e^{\zeta(z)\lambda} .
\end{equation}
This function is a solution of the Lam\`e equation
\begin{equation}
\Phi''(\lambda,z)=\Phi(\lambda,z) \bigl[ \wp(z)+2\wp(\lambda) \bigr] .
\end{equation}
From the monodromy properties of the Weierstrass functions it follows
that $\Phi(\lambda,z)$ is double-periodic as a function of $z$
though it is not elliptic in the classical sense due to the essential
singularity at $z=0$ for $\lambda\ne 0$. It also follows that
$\Phi(\lambda,z)$ is double-Bloch as a function of $\lambda$, namely
$$
\Phi(\lambda+2\omega_a,z)=T_a(z) \Phi(\lambda,z),\qquad
T_a(z)=\exp\left[ 2\omega_a\zeta(z)-2\eta_a z \right],\quad a=1,2.
$$
In the fundamental domain of the lattice defined by the periods $2\omega_1$,
$2\omega_2$ the function $\Phi(\lambda,z)$ has a unique pole at the point
$\lambda=0$ with the following expansion in the neighborhood of this point:
\begin{equation}\label{phil}
\Phi(\lambda,z)=\lambda^{-1}+O(\lambda) .
\end{equation}
Let $f(\lambda)$ be a double-Bloch function with Bloch multipliers~$B_a$.
The gauge transformation
$$
f(\lambda)\longmapsto \wt f(\lambda)=f(\lambda) e^{k\lambda}
$$
does not change the poles of $f$, and produces the double-Bloch
function~$\wt f(\lambda)$ with the Bloch multipliers
$\wt{B_a}=B_a e^{2k \omega_a}$.
The two pairs of Bloch multipliers~$B_a$ and~$\wt{B_a}$ connected by
such a relation are called equivalent. Note that for all the equivalent
pairs of Bloch multipliers the product $B_1^{\omega_2} B_2^{-\omega_1}$
is a constant depending only on the equivalence class.
Further note that any pair of Bloch multipliers may be represented in
the form
$$
B_a=T_a(z) e^{2\omega_a k},\qquad a=1,2,
$$
with an appropriate choice of the parameters $z$ and $k$.

There is no differentiation with respect to the variable~$\lambda$
in the equation (\ref{sch}). Thus, it will be sufficient to
study the double-Bloch solutions $\psi(x,t,\lambda)$ with
Bloch multipliers~$B_a$ such that $B_a=T_a(z)$ for some~$z$.

It follows from (\ref{phil}) that a double-Bloch function $f(\lambda)$
with simple poles $\lambda_i$ in the fundamental domain and with Bloch
multipliers $B_a=T_a(z)$  can be represented in the form
\begin{equation}\label{db}
f(\lambda)=\sum_{i=1}^N s_i \Phi(\lambda-\lambda_i,z),
\end{equation}
where $s_i$ is the residue of the function $f(\lambda)$ at the
pole~$\lambda_i$. Indeed, the difference of the left and right hand sides
in (\ref{db}) is a double-Bloch function with the same Bloch multipliers
as~$f(\lambda)$. It is also holomorphic in the fundamental domain.
Therefore, it equals zero since any non-zero double-Bloch
function with at least one of the Bloch multipliers distinct from~$1$
has at least one pole in the fundamental domain.

Now we are in position to present the generating problem for the
equations~(\ref{qyy}).
\begin{teo}
The equation \emph{(\ref{sch})} with the potential given by
\begin{equation}\label{ut1}
u(x,y,\lambda)=-2\sum_{i=1}^N \bigl[ (\lambda_{i\,x})^2\wp(\lambda-\lambda_i)
+\lambda_{i\,xx}\,\zeta(\lambda-\lambda_i) \bigr] +c(x,y),
\end{equation}
and the balanced set of poles \emph{(\ref{bal})}, has $N$ linearly independent
double-Bloch solutions with Bloch multipliers~$T_a(z)$,
that is, solutions of the form \emph{(\ref{db})}, if and only if
\begin{equation}\label{c}
c(x,y)=\frac{2}{Nh} U(q)-
\frac{1}{2Nh} \sum_{i=1}^N \frac{q_{i\,y}^2}{h-q_{i\,x}} ,
\end{equation}
and the functions $q_i(x,y)$ satisfy \emph{(\ref{qyy})}.

If \emph{(\ref{sch})} has $N$ linearly independent solutions of the form
\emph{(\ref{db})} for some $z$, then they exist for all values of $z$.
\end{teo}

\noindent{\it Proof.} We begin with a remark. In fact, if
$u(x,y,\lambda)$ is an elliptic function with a balanced set of
poles then it has to be of the form (\ref{ut1}) provided there
exist $N$ linearly independent double-Bloch solutions of
(\ref{sch}) for all values of the parameter $z$ in a neighborhood
of $z=0$.

Indeed, let us substitute (\ref{db}) into (\ref{sch}).
First of all, we conclude that the potential $u$ has at most double poles
at the points~$\lambda_i$. Thus, the potential is of the form
$$
u(\lambda,x,y)=\sum_{i=1}^N \bigl[
a_i \wp(\lambda-\lambda_i)+b_i \zeta(\lambda-\lambda_i) \bigr]+c(x,y)
$$
with some unknown coefficients $a_i=a_i(x,y)$ and $b_i=b_i(x,y)$.
Now, the coefficients of the singular part of the right hand side in~(\ref{sch})
must equal zero. The vanishing of the triple poles $(\lambda-\lambda_i)^{-3}$
implies $a_i=-2(\lambda_{i\,x})^2$.
The vanishing of the double poles $(\lambda-\lambda_i)^{-2}$ gives
the equalities
\begin{equation}\label{sx}
2s_{i\,x} \lambda_{i\,x}=s_i (\lambda_{i\,y}-\lambda_{i\,xx}-b_i)
-\sum_{j\ne i} s_j a_i \Phi(\lambda_i-\lambda_j,z) .
\end{equation}
Finally, the vanishing of the simple poles $(\lambda-\lambda_i)^{-1}$
leads to the equalities
\begin{equation}\label{sy}
\begin{aligned}
s_{i\,y}-s_{i\,xx}=s_i & \biggl( \lambda^2_{i\,x}\wp(z) +
\sum_{j\ne i} \bigl[ a_i \wp(\lambda_i-\lambda_j)+
b_j\zeta(\lambda_i-\lambda_j) \bigr] + c \biggr) +{}\\
{}&+\sum_{j\ne i} s_j \bigl( a_i \Phi'(\lambda_i-\lambda_j,z)+
b_j\Phi(\lambda_i-\lambda_j,z) \bigr) .
\end{aligned}
\end{equation}
The equations (\ref{sx}) and (\ref{sy}) are linear equations
for~$s_i=s_i(x,y,z)$. If we introduce the vector $\vS=(s_1,\dots,s_N)$
and the matrices $L=(L_{ij})$, $A=(A_{ij})$ with matrix elements
\begin{equation}\label{lij}
L_{ij}=\delta_{ij}\xi_i+(1-\delta_{ij}) \lambda_{i\,x}
\Phi(\lambda_i-\lambda_j,z),
\qquad\mbox{where}\qquad
\xi_i=\frac{\lambda_{i\,y}-\lambda_{i\,xx}-b_i}{2\lambda_{i\,x}},
\end{equation}
and
$$
\begin{aligned}
A_{ij} &=\delta_{ij}\biggl( \lambda^2_{i\,x}\wp(z) +\sum_{j\ne i}
\bigl[ -2\lambda_{i\,x}^2 \wp(\lambda_i-\lambda_j)+
b_j\zeta(\lambda_i-\lambda_j) \bigr] + c \biggr) +{}\\
{}&\qquad{}+(1-\delta_{ij}) \bigl( -2\lambda_{i\,x}^2
\Phi'(\lambda_i-\lambda_j,z)+ b_j\Phi(\lambda_i-\lambda_j,z) \bigr) .
\end{aligned}
$$
then the equations (\ref{sx}) and (\ref{sy}) can be written in the
form
\begin{equation}\label{la}
\vS_x=L \vS,\qquad \vS_y=\vS_{xx}+A \vS=(L^2+L_x+A)\vS.
\end{equation}
Let $M=L^2+L_x+A$, then the compatibility of the equations (\ref{la})
is equivalent to the zero-curvature equation for $L$ and $M$, i.e.
\begin{equation}\label{zc}
L_y-M_x+[L,M]=0 .
\end{equation}
The matrix elements of $M$ can be computed with the help of the
identities (\ref{pp}):
\begin{equation}\label{mij}
\begin{aligned}
M_{ii} &=\lambda_{i\,x} \biggl(\sum_{k=1}^N \lambda_{k\,x}\biggr)
\wp(z)+m_{i}^0, \\
M_{ij} &=-\lambda_{i\,x} \biggl(\sum_{k=1}^N \lambda_{k\,x}\biggr)
\Phi'(\lambda_i-\lambda_j,z)+m_{ij}\Phi(\lambda_i-\lambda_j,z),\qquad i\ne j
\end{aligned}
\end{equation}
where
$$
\begin{aligned}
m_{i}^0 &=\xi_i^2+\xi_{i\,x}-\sum_{k\ne i} \lambda_{k\,x}
\bigl( 2\lambda_{k\,x}^2+\lambda_{i\,x} \bigr)\wp(\lambda_i-\lambda_k)
+\sum_{k\ne i} b_k\zeta(\lambda_i-\lambda_k)+c ,\\
m_{ij} &=\lambda_{i\,x}(\xi_i+\xi_j)+\lambda_{i\,xx}+b_i+
\sum_{k\ne i,j} \lambda_{i\,x}\lambda_{k\,x}\,
\eta(\lambda_i,\lambda_k,\lambda_j) .
\end{aligned}
$$
The coefficients $b_i$ can be determined from the off-diagonal part
of the zero curvature equation. The left-hand side of the equation
corresponding to a pair of indexes $i\ne j$ is a double-periodic function
of~$z$. It is holomorphic except at $z=0$, where it has the form
$O(z^{-3})\exp\bigl[(\lambda_i-\lambda_j)\zeta(z)\bigr]$. Such a function equals
zero if and only if the corresponding coefficients at $z^{-3}$, $z^{-2}$
and $z^{-1}$ vanish. A direct computation shows that the coefficient
at $z^{-3}$ vanishes identically, while the coefficient at $z^{-2}$
equals
$$
\biggl( \sum_{k=1}^N \lambda_{k\,x} \biggr)(b_i+2\lambda_{i\,xx}) .
$$
Since our assumption prevents the first factor from vanishing we conclude
that $b_i=-2\lambda_{i\,xx}$. Given this, another direct computation shows
that the coefficient at~$z^{-1}$ also vanishes identically.

The zero-curvature equation (\ref{zc}) is not only a necessary but
also a sufficient condition for (\ref{sch}) to have solutions of
the form~(\ref{db}). The following lemma now completes the proof
of the theorem.
\begin{lem}
Let $L=(L_{ij}(x,y,z))$ and $M=(M_{ij}(x,y,z))$ be defined by the formulas
\emph{(\ref{lij})} and \emph{(\ref{mij})}, where
$b_i=-2\lambda_{i\,xx}$ and the set of $\lambda_i(x,y)$, $i=1,\dots, N$
is balanced. Then $L$ and $M$ satisfy the equation \emph{(\ref{zc})}
if and only if $c(x,y)$ is given by \emph{(\ref{c})} and the
functions $q_i(x,y)$ solve \emph{(\ref{qyy})}.
\end{lem}
\begin{proof}
It was mentioned above that all the off-diagonal equations in (\ref{zc}) become
identities if $b_i=-2\lambda_{i\,xx}$.
The diagonal part of the zero-curvature equation (\ref{zc}) simplifies with
the help of the identities (\ref{pp}) and (\ref{ppp}). Under a change
of variables $\lambda_i=q_i-h x$ it takes the form
\begin{equation}\label{qyyc}
\begin{aligned}
q_{i\,yy} &= -2(h-q_{i\,x})c_x+\biggl\{
\frac{q_{i\,xx}^2-q_{i\,y}^2}{h-q_{i\,x}} +q_{i\,xxx}
\biggr\}_{\!x}+{}\\
{}&\ {}+4(h-q_{i\,x}) \sum_{j\ne i} \bigl[ (h-q_{j\,x})^3 \wp'(q_i-q_j)
-3(h-q_{j\,x})q_{j\,xx}\,\wp(q_i-q_j)+q_{j\,xxx}\,\zeta(q_i-q_j) \bigr]
\end{aligned}
\end{equation}
Now consider the sum of the equations (\ref{qyyc}) for all $i$ from $1$ to~$N$.
Since the poles are balanced, the left hand side vanishes
and the coefficient at $c_x$ becomes $-2Nh$.
The other terms in the right hand side can be written as
$$
\frac{\p}{\p x} \biggl(
-\sum_{i=1}^N \frac{q_{i\,y}^2}{h-q_{i\,x}}+4U(q) \biggr) .
$$
Therefore, $c$ is given by (\ref{c}) up to an arbitrary function of $y$, which
does not affect the equations (\ref{qyyc}). Finally, substituting (\ref{c})
into (\ref{qyyc}) we arrive at~(\ref{qyy}).
\end{proof}

\section{Field analog of the elliptic Calogero\,--\,Moser system}

In this section we show that equations (\ref{qyy}) can be obtained
as a reduction of the field elliptic CM system.

In \cite{n} the elliptic CM system was identified with a particular case
of the Hitchin system on an elliptic curve with a puncture. In \cite{krvb}
a Hamiltonian theory of zero-curvature equations on algebraic curves was
developed and identified with infinite-dimentional field analogs of the
Hitchin system.
In particular, it was shown that the zero-curvature equation on an
elliptic curve with a puncture can be seen as a field generalization
of the elliptic~CM system.

The field elliptic CM system is a Hamiltonian system on the space of
functions $\{q_i(x),p_i(x)\}_{i=1}^N$ with the canonical Poisson brackets
$$
\bigl\{ q_i(x),q_j(\tilde x)\bigr\}=\bigl\{ p_i(x),p_j(\tilde
x)\bigr\}=0, \quad \bigl\{ q_i(x),p_j(\tilde x)
\bigr\}=\delta_{ij}\,\delta (x-\tilde x) , \qquad 1\le i,j\le N ,
$$
Its Hamiltonian is given by (\ref{ham}). Note, that $\wt U(q)$ is elliptic
function of each of the variables $q_i$, $i=1,\dots,N$.
Substituting the definition of $\wt U(q)$ into (\ref{ham}) we obtain
the following expression for the hamiltonian density:
$$
\begin{aligned}
H &= \sum_{i=1}^N p_i^2(h-q_{i\,x})-
 \frac{1}{N h} \biggl( \sum_{i=1}^N p_i(h-q_{i\,x}) \biggr)^{\!2} -{}\\
{}&\quad{}-\sum_{i=1}^N \frac{q^2_{i\,xx}}{4(h-q_{i\,x})}
 -\frac{1}{2} \sum_{i\neq j}
 \bigl[ q_{i\,x}q_{j\,xx}-q_{j\,x}q_{i\,xx} \bigr] \zeta(q_i-q_j)+{}\\
{}&\quad{}+\frac{1}{2}\sum_{i\neq j}
 \bigl[(h-q_{i\,x})^2(h-q_{j\,x})+(h-q_{i\,x})(h-q_{j\,x})^2
 -h(q_{i\,x}-q_{j\,x})^2 \bigr]\wp(q_i-q_j) .
\end{aligned}
$$
The equations of motion are
\begin{equation}\label{sys}
\begin{aligned}
\dot q_i &=2p_i(h-q_{i\,x})-\frac{2}{Nh}
 \sum_{k=1}^N p_k (h-q_{k\,x})(h-q_{i\,x}) , \\
\dot p_i &=-2p_ip_{i\,x}+
 \frac{2}{Nh}\biggl\{\sum_{k=1}^N p_i p_k(h-q_{k\,x})\biggr\}_{\!x}+
 \biggl\{
 \frac{q_{i\,xxx}}{2(h-q_{i\,x})}+\frac{q_{i\,xx}^2}{4(h-q_{i\,x})^2}
 \biggr\}_{\!x}+{}\\
{}&\quad{}+2\sum_{j\ne i}\bigl[
 q_{j\,xxx}\zeta(q_i-q_j)-3(h-q_{j\,x})q_{j\,xx}\wp(q_i-q_j)+
 (h-q_{j\,x})^3\wp'(q_i-q_j) \bigr]
\end{aligned}
\end{equation}

Let us make a remark on the notations. Throughout this section by dots
we mean derivatives with respect to the variable~$y$, which we treat as
a time variable.
In view of the connection with the KP equation this time variable
corresponds to the second time of the KP hierarchy, for which~$y$ is
a standard notation.

It is easy to check that the subspace $\cal N$ defined by the constraint
\begin{equation}\label{con}
\sum_{i=1}^N q_i(x)=const ,
\end{equation}
is invariant for the system (\ref{sys}).
On that subspace the first two terms of the Hamiltonian density~$H$ can be
represented in the form
\begin{equation}
H=\frac{1}{2Nh}\biggl(\sum_{i\ne j} (p_i-p_j)^2 (h-q_{i\,x})(h-q_{j\,x})
\biggr) -\wt U(q) .
\end{equation}
Therefore, the Hamiltonian (\ref{ham}) restricted to ${\cal N}$ is invariant
under the transformation
\begin{equation}\label{psym}
p_i(x)\to p_i(x)+f(x),
\end{equation}
where $f(x)$ is an arbitrary function.
The constraint (\ref{con}) is the Hamiltonian of that symmetry. The canonical
symplectic form is also invariant with respect to (\ref{con}). Therefore, the
Hamiltonian system (\ref{sys}) restricted to $\cal N$ can be reduced to a
factor space. The reduction can be described as follows.

Let us define the variables $\ell_i=p_i+\kappa$, $i=1,\dots,N$, where
\begin{equation}\label{kap}
\kappa=-\frac{1}{Nh}\sum_{k=1}^N p_k(h-q_{k\,x}).
\end{equation}
They are invariant with respect to the symmetry (\ref{psym}), and satisfy
the equation
\begin{equation}\label{conl}
\sum_{k=1}^N \ell_k(h-q_{k\,x})=0.
\end{equation}
A direct substitution shows that equations (\ref{sys}) imply the system
of equation
\begin{equation}\label{sys1}
\begin{aligned}
\dot q_i&=2\ell_i (h-q_{ix}) ,\\
\dot \ell_i&=-2\ell_i\ell_{i\,x}+\frac{2}{Nh} \biggl\{ \sum_{k=1}^N
 \ell_k^2(h-q_{kx}) - U(q) \biggr\}_{\!x} + \biggl\{
 \frac{q_{i\,xxx}}{2(h-q_{i\,x})}+\frac{q_{i\,xx}^2}{4(h-q_{i\,x})^2}
\biggr\}_{\!x} + {}\\
{}&\qquad{}+2 \sum_{j\ne i} \bigl[ (h-q_{j\,x})^3 \wp'(q_i-q_j)
 -3(h-q_{j\,x})q_{j\,xx}\,\wp(q_i-q_j)+q_{j\,xxx}\,\zeta(q_i-q_j) \bigr] ,
\end{aligned}
\end{equation}
\begin{teo}
Equations \emph{(\ref{qyy})} are equivalent to the restriction of
the system \emph{(\ref{sys1})} to the subspace $\cal M$ defined
by the constraints \emph{(\ref{con})} and \emph{(\ref{conl})}.
\end{teo}
\begin{proof}
Let us show that equations (\ref{qyy}) imply (\ref{sys1}).
The first equations can be regarded as the definition of $\ell_i$,
$i=1,\dots,N$. Taking their derivative we obtain
\begin{equation}\label{ddq}
\ddot q_i=2\dot\ell_i(h-q_{i\,x})-2\ell_i
\bigl( 2\ell_{i\,x}(h-q_{i\,x})-2\ell_i q_{i\,xx}\bigr) .
\end{equation}
Therefore,
$$
\dot\ell_i=2\ell_i\ell_{i\,x}-2\ell_i^2 \frac{q_{i\,xx}}{h-q_{i\,x}}
+\frac{\ddot q_i}{2(h-q_{i\,x})} .
$$
To obtain the second equations of (\ref{sys1}) it is now sufficient
to substitute the right hand side of (\ref{qyyc}) for $\ddot q_i$
and use formula (\ref{c}).

The equation (\ref{ddq}) can also be used to derive (\ref{qyy})
from (\ref{sys1}) in a straightforward manner.
\end{proof}

Note that a solution of (\ref{sys1}) restricted to the the subspace $\cal M$
defines a solution of (\ref{sys}) uniquely up to initial data. Namely,
it can be checked directly that if $\kappa(x,y)$ as a solution of the equation
\begin{equation}\label{kap}
\dot \kappa= \biggl\{-\kappa^2 +
\frac{2}{Nh}\sum_{k=1}^N \ell_i^2(h-q_{kx})-\frac{2}{Nh} U(q)
\biggr\}_{\!x}\ ,
\end{equation}
and $\ell_i$, $q_i$ is a solution of (\ref{sys1}) on $\cal M$, then
$q_i$, $p_i=\ell_i-\kappa$ is a solution of (\ref{sys}).

Our final goal for this section is to present the Lax pair for the
field elliptic CM system.
\begin{teo}
System \emph{(\ref{sys})} admits the zero-curvature representation,
i.\,e.~it is equivalent to the matrix equation
$$
\wt L_y-\wt M_x+\bigl[\wt L, \wt M \bigr]=0,
$$
with the Lax matrices $\wt L=\bigl(\wt L_{ij}\bigr)$ and
$\wt M=\bigl(\wt M_{ij}\bigr)$ of the form
\begin{equation}\label{lax}
\begin{aligned}
\wt L_{ij}&=-\delta_{ij}p_i+(1-\delta_{ij})\alpha_i\alpha_j\Phi(q_i-q_j,z), \\
\wt M_{ij}&=\delta_{ij} \bigl[ -Nh \alpha_i^2 \wp(z)+ \wt m_{i}^0 \bigr]+
(1-\delta_{ij}) \alpha_i\alpha_j
\bigl[ Nh\,\Phi'(q_i-q_j,z)- \wt m_{ij} \Phi(q_i-q_j,z) \bigr] ,
\end{aligned}
\end{equation}
where $\alpha_i^2=q_{i\,x}-h$,
$$
\begin{aligned}
\wt m_i^0 &=p_i^2+\frac{\alpha_{ixx}}{\alpha_i}+2\kappa p_i-
\sum_{j\neq i} \bigl[ \alpha_j^2(2\alpha_i^4+\alpha_j^2)\wp(q_i-q_j)+
4\alpha_i\alpha_{i\,x}\zeta(q_i-q_j) \bigr],\\
\wt m_{ij}&=p_i+p_j+2\kappa+\frac{\alpha_{i\,x}}{\alpha_i}-
\frac{\alpha_{j\,x}}{\alpha_j}
+\sum_{k\neq i,j}\alpha_k^2 \eta(q_i,q_k,q_j),
\end{aligned}
$$
and $\kappa$ is given by \emph{(\ref{kap})}.
\end{teo}
\begin{proof}
If we apply to the matrices $L$ and $M$ given by
(\ref{lij}) and (\ref{mij}) a gauge transformation
$$
L\longmapsto g_x g^{-1}+g L g^{-1},\qquad\quad
M\longmapsto g_y g^{-1}+g M g^{-1},
$$
where~$g$ is a diagonal matrix,
$g=(g_{ij})$, $g_{ij}=\delta_{ij}(\lambda_{i\,x})^{-1/2}$,
and then substitute
$\lambda_i=q_i-h x$ and $\lambda_{i\,y}/2\lambda_{i\,x}=\ell_i$,
that would give us a Lax pair for system (\ref{sys1}).
To obtain (\ref{lax}) we apply another gauge tranformation with
$g=e^{K} I$ and substitute $\ell_i=p_i+\kappa$, $i=1,\dots,N$.
Here $K=K(x,y)=\int^{x} \kappa(\wt x,y)\,d{\wt x}$. Note that
$K_y=-\kappa^2-c$ due to (\ref{kap}) and (\ref{c}).
\end{proof}

\section{Elliptic families of solutions of the KP equation}

The KP equation (\ref{kp}) is equivalent to the commutation condition
\begin{equation}\label{lakp}
\left[ \p_y-\cL, \p_t-\cA\right]=0,
\qquad \p_y=\p/\p y,\quad \p_t=\p/\p t,
\end{equation}
for the auxiliary linear differential operators
$$
\cL=\p^2_{xx}+u(x,y,t), \qquad
\cA=\p^3_{xxx}+\frac{3}{2} u \p_x+w(x,y,t),\qquad
\p_x=\p/\p x.
$$
We use this representation in order to derive our main result.
\begin{teo}
Let $u(x,y,t,\lambda)$ be an elliptic family of solution to
the KP equation that has a balanced set of poles
$\lambda_i(x,y,t)=q_i(x,y,t)-h x$, $i=1,\dots,N$.
Then $u(x,y,t,\lambda)$ has the form \emph{(\ref{ukp})} and the dynamics
of the functions $q_i(x,y,t)$ with respect to~$y$ is described by
the system~\emph{(\ref{qyy})}.
\end{teo}
\begin{proof}
Substituting $u$ into (\ref{kp}) we immediately conclude that $u$ may have poles
in~$\lambda$ of at most second order. Moreover, comparing the coefficients
of the expansions of the left and right hand sides in (\ref{kp}) near the
pole~$\lambda_i$, we deduce that the principal part of the solution~$u$ coincides
with the one given by (\ref{ukp}).

Next step is to show that operator equation (\ref{lakp}) implies
the existence of double-Bloch solutions for the equation
$(\p_y-\cL)\psi(x,y,t,\lambda)=0$.

Let us define a matrix $S(x,y,t,z)$ to be a solution of the linear differential
equation $\p_x S=L S$, where $L=(L_{ij})$,
$$
L_{ij} =\delta_{ij}\left(
\frac{\lambda_{i\,y}+\lambda_{i\,xx}}{2\lambda_{i\,x}}
\right)+(1-\delta_{ij}) \lambda_{i\,x}\, \Phi(\lambda_i-\lambda_j,z) ,
$$
with the initial conditions $S(0,y,t,z)=S_0(y,t,z)$, a non-singular matrix.
By $\Phi$ we denote the row-vector
$\bigl( \Phi(\lambda-\lambda_1,z),\dots,\Phi(\lambda-\lambda_N,z) \bigr)$.
It follows immediately that the vector $(\p_y-\cL)\Phi S$ has at most simple
poles at~$\lambda_i$, $i=1,\dots,N$. Therefore, it is equal to $\Phi D$ for
some matrix~$D$. The commutation relation (\ref{lakp}) implies that
$D_x= L D$. To show this, consider the vector
$$
(\p_t-\cA)\Phi D=(\p_t-\cA)(\p_t-\cL)\Phi S=(\p_y-\cL)(\p_t-\cA)\Phi S.
$$
It has the poles of the at most third order and therefore the vector
$(\p_t-\cA)\phi S$ has at most simple poles. In this case, however, the vector
$$
(\p_t-\cA)(\p_t-\cL)\Phi S=(\p_y-\cL)(\p_t-\cA)\Phi S=(\p_t-\cA)\Phi D
$$
has the poles of the at most second order. Vanishing of the poles of the third order
in the expression $(\p_t-\cA)\Phi D$ is equivalent to the equation $D_x=LD$.

Since $S$ and $D$ are solutions to the same linear differential equation in $x$ they
differ by an $x$-independent matrix, namely $D(x,y,t,z)=S(x,y,t,z)T(y,t,z)$.
Let us define a matrix $F(y,t,z)$ from the equation $\p_y F+T F=0$ and the
initial condition $F(0,t,z)=I$. Here $I$ is the identity matrix.
Let $\wt S=S F$, then
$$
(\p_y-\cL)\phi\wt S=(\p_y-\cL)\phi S F=\phi D F+\phi S F_y=
\phi S\left(T F+F_y\right)=0,
$$
and the components of the vector $\phi\wt S$ are independent double-Bloch solutions
to (\ref{sch}).

To conclude the proof it now suffices to apply Theorem~$1$.
\end{proof}

\section{The algebraic-geometric solutions}

According to \cite{kr}, a smooth genus $g$ algebraic curve
$\Gamma$ with fixed local coordinate $w$ at a puncture $P_0$ defines solutions
of the entire KP hierarchy by the formula
$$
u(t)=2\frac{\p^2}{\p x^2} \ln \theta
\left(\biggl. \sum\nolimits_k \vU_k t_k + \vZ\ \bigr|\ B\right)+const .
$$
Here $B=(B_{jk})$ is a matrix of the $b$-periods of normalized holomorphic
differentials $\omega^h_k$
\begin{equation}\label{hd}
\oint_{a_i}\omega^h_j=\delta_{ij},\qquad\qquad
B_{ij}=\oint_{b_i}\omega^h_j,
\end{equation}
while the vectors $\vU_k=(\vU_k^j)$ are vectors of the $b$-periods
$$
\vU_k^{j}=\frac{1}{2\pi i}\oint_{b_j}d\Omega_k,\qquad\qquad
\oint_{a_j}d\Omega_k=0,
$$
of the normalized meromorphic differentials of the second kind~$d\Omega_k$,
defined by their expansions
\begin{equation}\label{om}
d\Omega_k=dw^{-k}+O(1)dw
\end{equation}
in the neighborhood of~$P_0$.

Let $\Gamma$ be a $N$-fold branched cover of an elliptic curve~$\cE$:
$$
\rho\colon\Gamma\longrightarrow\cE .
$$
Then the induced map of the Jacobians defines an embedding of
$\cal E$ into $J(\Gamma)$, i.e. $\rho^*\cE\subset J(\Gamma)$.
Therefore, any $N$-fold cover of $\cal E$ defines an elliptic family
of solutions of the KP equation. The following assertion shows that
the corresponding solutions have exactly $N$ poles.
Moreover, if the local coordinate~$w$ at
the puncture~$P_0$ is $\rho^*(\lambda)$, then the poles are balanced.
Here $\lambda$ is a flat coordinate on $\cE$.

\begin{teo}
Let $\Gamma$ be a smooth $N$-fold branched cover of the elliptic
curve~$\cE$, and let $P_0\in\Gamma$ be a preimage of the point~$\lambda=0$
on~$\cE$. Let $d\Omega_k$ be a normalized meromorphic differential
on~$\Gamma$ with the only pole at~$P_0$ of the form \emph{(\ref{om})},
where $w=\rho^{*}(\lambda)$, and let $2\pi i \vU$ and $2\pi i \vV$ be the
vectors of $b$-periods of the differentials~$d\Omega_1$
and~$d\Omega_2$ respectively. Then the equation
\begin{equation}
\theta \bigl( \vL\lambda+\vU x+\vV y\ |\ B \bigr)=0
\end{equation}
has~$N$ balanced roots $\lambda_i(x,y)=q_i(x,y)-x/N$, $\sum_i q_i(x,y)=0$,
and the functions $q_i$ satisfy system~\emph{(\ref{qyy})}.
\end{teo}
\begin{proof}
Let $2\omega_1$, $2\omega_2$ be the  periods of~$\cE$,
such that $\Im(\tau)=\Im(\omega_2/\omega_1)>0$.
The Jacobian $J(\Gamma)$ is the factor of~$\comp^g$ over the lattice~$\cB$,
spanned by the basis vectors $\Vec{e}_i\in\comp^g$, $i=1,\dots,g$
and the columns $\vB_i=(B_{ij})\in\comp^g$, $i=1,\dots,g$, of the matrix~$B$.
Let $\vL$ be a vector in $\comp^g$ that spans $\rho^*\cE\subset J(\Gamma)$.
Note that not only $\vL\in\cB$, but also $\tau\vL\in\cB$.

The function
$\theta \bigl( \sum _k \vU_k t_k+\vL \lambda+\vZ\ |\ B\bigr)$
as a function of~$\lambda$ has a finite number~$D$ of zeros.
Its monodromy properties (\ref{thm}) imply that it can be written as
$$
\theta \left(\sum\nolimits_k \vU_k t_k +\vL\lambda+\vZ\ \bigr|\ B\right)=
f(t)e^{c_1\lambda+c_2\lambda^2}
\prod_{i=1}^D \sigma \bigl( \lambda-\lambda_i(t) \bigr) ,
$$
where $c_1$, $c_2$ are constants.

Note, that the $\lambda_i$'s are defined modulo the periods of~$\cE$.
In order to count them we integrate $d\ln \theta$ along the
boundary of the fundamental domain of $\rho^*\cE$ in~$\comp^g$.

The embedding of $\cE$ in $J(\Gamma)$ is defined by equivalence classes
of the divisors $\rho^*(z)-\rho^*(0)$, where $\rho^*(z)$ is the divisor of
preimages on $\Gamma$ of a point $z\in\cE$.
Preimages on $\Gamma$ of $a$ and $b$-cycles of $\cal E$ are some
linear combination of the basis cycles on $\Gamma$, i.e.
$$
\rho^* a=\sum_{k=1}^g n_ka_k + m_k b_k,\qquad
\rho^* b=\sum_{k=1}^g n_k'a_k+ m_k'b_k .
$$
Therefore, the vector $\vL$ equals
$$
\vL=\sum_{k=1}^g n_k \Vec{e}_k+m_k \vB_k,\qquad
\tau \vL=\sum_{k=1}^g n_k' \Vec{e}_k+m_k' \vB_k .
$$
The usual residue arguments imply
$$
2\pi i D=\oint_{\p (\rho^*\cE)} d\ln \theta=
\int_{\tau \vL}\left(\int_{\vL}d\ln\theta\right)-
\int_{\vL}\left(\int_{\tau \vL}d\ln\theta\right)
$$
The monodromy properties of the theta-function imply
$$
D=\sum_{k=1}^g \bigl( n_k m_k'-n_k' m_k \bigr).
$$
The right hand side in the last formula is the intersection number of
the cycles~$\rho^{*}a$ and~$\rho^{*}b$, i.\,e.
$$
D=(\rho^*a)\cap (\rho^*b)=N \left(a\cap b\right)=N,
$$
so the the theta-function has exactly~$N$ zeros $\lambda_i$, $i=1,\dots,N$.

Now let us show that the set of $\lambda_i$'s is balanced.
In a way similar to the residue argument above we find
\begin{equation}\label{plpt}
-2\pi i\sum_{j=1}^N \frac{\p \lambda_j}{\p t_k}=
\oint_{\p (\rho^*\cE)} \left(\p_{t_k}\ln \theta\right) d\lambda=
\int_{b}d\lambda\left(\int_{\rho^*a}d\Omega_k\right)-
\int_{a}d\lambda\left(\int_{\rho^*b}d\Omega_k\right)
\end{equation}
Let $\Tr d\Omega=\rho_*(d\Omega_k)$ be the sum of~$d\Omega_k$ on all the
sheets of~$\Gamma$ over a point $\lambda\in \cE$. It is a meromorphic
differential on~$\cE$. Since the local coordinate~$w$ near the puncture
is defined by the projection~$\rho$ we have
\begin{equation}
\Tr d\Omega_k=\frac{(-1)^{k}}{(k-1)!} \wp^{(k-1)}(\lambda)\,d\lambda
+r_k d\lambda,
\end{equation}
where $r_k$ is a constant.
The right hand side in (\ref{plpt}) can be written as
$2\pi i \res_{\lambda=0} (\Tr \Omega_k)\,d\lambda$.
When $k>1$ it equals zero, while for $k=1$ we have
$$
\res_{\lambda=0} (\Tr\Omega_1)\,d\lambda=
\res_{\lambda=0} \zeta(\lambda)\,d\lambda=1 .
$$
Therefore, we obtain
\begin{equation}
\sum_{i=1}^N \frac{\p\lambda_j}{\p x}=-1,\qquad\qquad
\sum_{i=1}^N \frac{\p\lambda_j}{\p t_k}=0,\quad k>1 .
\end{equation}
and consequently the set $\lambda_i$, $i=1,\dots, N$ satisfy (\ref{bal}).
Note, that our choice of a local coordinate near the puncture corresponds
to $h=1/N$. An arbitrary non-zero value of~$h$ may be obtained by setting
$w=\rho^{*}(\lambda/N h)$.
Theorem~5 is proved.
\end{proof}

\begin{rem}
If $q_i(x,y)$, $i=1,\dots,N$ are periodic functions of~$x$,
then the algebraic curve~$\Gamma$ can be identified with the spectral
curve for the equation $(\p_x- L)\vS=0$ (see \cite{krvb}).
\end{rem}

\appendix
\section{Appendix}
\subsection{Elliptic functions}

Here we list the definitions and basic properties of the classical
elliptic functions (see \cite{bat} for details).
Let $2\omega_1,2\omega_2\in\comp$
be a pair of periods, $\Im (\omega_2/\omega_1)>0$.
The Weierstrass sigma-function is defined by the infinite product
$$
\sigma(z)= z \prod\nolimits_{m^2+n^2\ne 0} \left( 1-\frac{z}{\omega_{m n}}
\right)
\exp\left\{ \frac{z}{\omega_{m n}}+\frac{z^2}{2 \omega^2_{m n}} \right\},
\qquad
\omega_{m n}=2m \omega_1+2n \omega_2 .
$$
The product converges for every $z$ to an entire function with simple
zeros at the points $z=\omega_{m n}$.
The Weierstrass zeta-function and $\wp$-function are then defined by
$$
\zeta(z)=\frac{\sigma'(z)}{\sigma(z)},\qquad
\wp(z)=-\zeta'(z) .
$$
It follows directly from this definition that $\sigma(z)$ and $\zeta(z)$
are odd functions while $\wp(z)$ is an even function. Under shifts
of the periods the Weierstrass functions transform as follows:
$$
\sigma(z+2\omega_a)=e^{2\eta_a(z+\omega_a)}\sigma(z),\quad
\zeta(z+2\omega_a)=\zeta(z)+2\eta_a,\qquad a=1,2,
$$
where $\eta_a=\zeta(\omega_a)$ and $\eta_1\omega_2-\eta_2\omega_1=\pi i/2$.
The $\wp$-function is double-periodic
$$
\wp(z+2\omega_1)=\wp(z+2\omega_2)=\wp(z)=\wp(-z)
$$
and can be regarded as a function on the elliptic curve
$\Gamma=\comp \bigl/ \inte[2\omega_1,2\omega_2] \bigr.$
where it has the only (double) pole at $z=0$.
It is useful to write the Laurent expansions of the Weierstrass
functions in the neighborhood of $z=0$:
$$
\sigma(z)=z+O(z^5),\qquad \zeta(z)=\frac{1}{z}+O(z^3),\qquad
\wp(z)=\frac{1}{z^2}+O(z^2) .
$$

\subsection{Identities with the function $\Phi(\lambda,z)$}

Here we collect some useful identities involving the function
$\Phi(\lambda,z)$, which is defined by (\ref{phi}).

The derivative of the function $\Phi(\lambda,z)$ with respect to the
variable $\lambda$ equals
\begin{equation}
\Phi'(\lambda,z) = \Phi(\lambda,z)
\bigl[ \zeta(z)-\zeta(\lambda)-\zeta(z-\lambda) \bigr]
\end{equation}
We also have the following product identities:
\begin{equation}\label{pp}
\begin{aligned}
\Phi(\lambda-\mu,z)\Phi(\mu-\lambda,z) &=\wp(z)-\wp(\lambda-\mu),\\
\Phi(\lambda-\nu,z)\Phi(\nu-\mu,z) &= -\Phi'(\lambda-\mu,z)+
\Phi(\lambda-\mu,z) \eta(\lambda,\nu,\mu) ,
\end{aligned}
\end{equation}
where in the second equation we use the notation
$$
\eta(\lambda,\nu,\mu)=
\zeta(\lambda-\nu)+\zeta(\nu-\mu)-\zeta(\lambda-\mu) .
$$
Note that $\eta$ is a completely antisymmetric function of its
arguments.
To complete the list of the identities required for our
computations we differentiate formulas (\ref{pp}) to get
\begin{equation}\label{ppp}
\begin{aligned}
\Phi'(\lambda-\mu,z)\Phi(\mu-\lambda,z)-
\Phi(\lambda-\mu,z)\Phi'(\mu-\lambda,z) &=-\wp'(\lambda-\mu),\\
\Phi'(\lambda-\nu,z)\Phi(\nu-\mu,z)-
\Phi(\lambda-\nu,z)\Phi'(\nu-\mu,z) &=
-\Phi(\lambda-\mu) \bigl[ \wp(\lambda-\nu)-\wp(\nu-\mu) \bigr] .
\end{aligned}
\end{equation}

\subsection{Riemann $\theta$-function}

Let $\Gamma$ be a genus $g$ algebraic curve with fixed basis of cycles
$a_i$, $b_i$, $i\le 1\le g$ with intersections $a_i\circ b_j=\delta_{ij}$.
Let~$B$ be the matrix of normalized holomorphic differentials~$\omega^h_i$,
see (\ref{hd}). Then~$B$ is a Riemann matrix, i.e. a symmetric $g\times g$
matrix with positive definite imaginary part $\Im B<0$.

The Riemann $\theta$-function, associated with the curve $\Gamma$ is an
analitic function of $g$ complex variables $\Vec{z}=(z_1,\dots,z_g)$,
defined by its Fourier expansion
$$
\theta(\Vec{z}\,|\,B)=\sum\nolimits_{\Vec{n}\in\inte^g}
\ e^{2\pi i (\Vec{m},\Vec{z}) + \pi i (B\Vec{m},\Vec{m})}\ .
$$
The Riemann $\theta$-function has the following monodromy properties
with respect to the lattice~$\cB$, spanned by the basis vectors
Let $\Vec{e}_i\in\comp^g$, $i=1,\dots,g$ and the columns $B_i\in\comp^g$
of the matrix~$B$:
\begin{equation}\label{thm}
\begin{aligned}
\theta(\Vec{z}+\Vec{n}\,|\,B) &=\theta(\Vec{z}\,|\,B),\\
\theta(\Vec{z}+B\Vec{n}\,|\,B) &=
\exp\bigl[ -2\pi i(\Vec{n},\Vec{z})-\pi i (B\Vec{n},\Vec{n}) \bigr]
\theta(\Vec{z}\,|\,B) .
\end{aligned}
\end{equation}
Here $\Vec{n}$ is a vector with integer components.


\end{document}